\documentclass[twocolumn]{aastex62}
\graphicspath{{./}{figures/}}
\usepackage{verbatim}
\shorttitle{Deep FRB Search of SGRB Sites}
\shortauthors{Madison et al.}
\begin{document}

\title{A Deep Targeted Search for Fast Radio Bursts from the Sites of Low-Redshift Short Gamma-Ray Bursts}
\correspondingauthor{D. R. Madison}
\email{dustin.madison@mail.wvu.edu}

\author[0000-0003-2285-0404]{D. R. Madison}
\affil{Department of Physics and Astronomy, West Virginia University, P.O. Box 6315, Morgantown, WV 26506, USA}
\affil{Center for Gravitational Waves and Cosmology, West Virginia University, Chestnut Ridge Research Bldg, Morgantown, WV 26505, USA}

\author[0000-0003-0385-491X]{D. Agarwal}
\affil{Department of Physics and Astronomy, West Virginia University, P.O. Box 6315, Morgantown, WV 26506, USA}
\affil{Center for Gravitational Waves and Cosmology, West Virginia University, Chestnut Ridge Research Bldg, Morgantown, WV 26505, USA}

\author[0000-0002-2059-0525]{K. Aggarwal}
\affil{Department of Physics and Astronomy, West Virginia University, P.O. Box 6315, Morgantown, WV 26506, USA}
\affil{Center for Gravitational Waves and Cosmology, West Virginia University, Chestnut Ridge Research Bldg, Morgantown, WV 26505, USA}

\author{O. Young}
\affil{Department of Physics and Astronomy, West Virginia University, P.O. Box 6315, Morgantown, WV 26506, USA}
\affil{Center for Gravitational Waves and Cosmology, West Virginia University, Chestnut Ridge Research Bldg, Morgantown, WV 26505, USA}

\author[0000-0002-6039-692X]{H. T. Cromartie}
\affil{Department of Astronomy, University of Virginia, 530 McCormick Rd., Charlottesville, VA 22903, USA}

\author[0000-0003-0721-651X]{M. T. Lam}
\affil{School of Physics and Astronomy, Rochester Institute of Technology, Rochester, NY 14623, USA}
\affil{Department of Physics and Astronomy, West Virginia University, P.O. Box 6315, Morgantown, WV 26506, USA}
\affil{Center for Gravitational Waves and Cosmology, West Virginia University, Chestnut Ridge Research Bldg, Morgantown, WV 26505, USA}

\author[0000-0002-2878-1502]{S. Chatterjee}
\affil{Department of Astronomy, Cornell University, 616-A Space Sciences Building, Ithaca, NY 14853, USA}
\affil{Cornell Center for Astrophysics and Planetary Science, 104 Space Sciences Building, Ithaca, NY 14853, USA}

\author{J. M. Cordes}
\affil{Department of Astronomy, Cornell University, 616-A Space Sciences Building, Ithaca, NY 14853, USA}
\affil{Cornell Center for Astrophysics and Planetary Science, 104 Space Sciences Building, Ithaca, NY 14853, USA}

\author{N. Garver-Daniels}
\affil{Department of Physics and Astronomy, West Virginia University, P.O. Box 6315, Morgantown, WV 26506, USA}
\affil{Center for Gravitational Waves and Cosmology, West Virginia University, Chestnut Ridge Research Bldg, Morgantown, WV 26505, USA}

\author{D. R. Lorimer}
\affil{Department of Physics and Astronomy, West Virginia University, P.O. Box 6315, Morgantown, WV 26506, USA}
\affil{Center for Gravitational Waves and Cosmology, West Virginia University, Chestnut Ridge Research Bldg, Morgantown, WV 26505, USA}

\author{R. S. Lynch}
\affil{Green Bank Observatory, P.O. Box 2, Green Bank, WV 24944, USA}

\author[0000-0001-7697-7422]{M. A. McLaughlin}
\affil{Department of Physics and Astronomy, West Virginia University, P.O. Box 6315, Morgantown, WV 26506, USA}
\affil{Center for Gravitational Waves and Cosmology, West Virginia University, Chestnut Ridge Research Bldg, Morgantown, WV 26505, USA}

\author[0000-0001-5799-9714]{S. M. Ransom}
\affil{The National Radio Astronomy Observatory, 520 Edgemont Rd., Charlottesville, VA, 22903, USA}

\author{R. S. Wharton}
\affil{Max-Planck-Institut f\"ur Radioastronomie, Auf dem H\"ugel 69, D-53121 Bonn, Germany}

\begin{abstract}
Some short gamma-ray bursts (SGRBs) are thought to be caused by the mergers of binary neutron stars which may sometimes produce massive neutron star remnants capable of producing extragalactic fast radio bursts (FRBs). We conducted a deep search for FRBs from the sites of six low-redshift SGRBs. We collected high time- and frequency-resolution data from each of the sites for 10 hours using the 2 GHz receiver of the Green Bank Telescope. Two of the SGRB sites we targeted were visible with the Arecibo Radio Telescope with which we conducted an additional 10 hours of 1.4 GHz observations for each. We searched our data for FRBs using the GPU-optimized dedispersion algorithm \textsc{heimdall} and the machine-learning-based package \texttt{FETCH} (Fast Extragalactic Transient Candidate Hunter). We did not discover any FRBs, but would have detected any with peak flux densities in excess of 87~mJy at the Green Bank Telescope or 21~mJy at Arecibo with a signal-to-noise ratio of at least 10. The isotropic-equivalent energy of any FRBs emitted from these sites in our bands during our observations must not have exceeded a few times $10^{38}$~erg, comparable to some of the lowest energy bursts yet seen from the first known repeating FRB 121102. 
\end{abstract}

\section{Introduction} \label{sec:intro}
With the detection of GW170817, the Laser Interferometer Gravitational-Wave Observatory (LIGO) linked a double neutron star merger to a short gamma-ray burst \citep[SGRB;][]{aaa+17_a,aaa+17_b}. Such a connection had been suspected for more than a decade in advance of that discovery. The distribution of gamma-ray burst (GRB) durations is clearly bimodal \citep{kmf+93}. Long GRBs, those lasting longer than approximately two seconds, are commonly found in star-forming regions of star-forming galaxies and are often accompanied by Type Ic supernovae. So-called ``collapsars", fireballs generated by rapid accretion onto black holes formed in the core collapse of massive, rapidly rotating stars, are the commonly accepted source of long GRBs \citep{w93}. On the other hand, SGRBs with durations less than two seconds---sometimes as short as tens of milliseconds---are not associated with active star-forming regions but are instead found in galaxies with ancient stellar populations. No supernovae are found to accompany SGRBs. The double neutron star merger model for SGRB progenitors---or potentially neutron star-black hole mergers---parsimoniously explains the short timescale for the gamma-ray emission, the lack of supernova association, the age of the stellar population in the host galaxies, and through natal kicks imparted to the neutron stars, the large spatial separation of SGRBs from star-forming regions \citep[see][and references therein]{b14}.

Some SGRBs have shown persistent X-ray emission tens of seconds to weeks after the initial gamma-ray flash that is difficult to explain with conventional models for the afterglow of an SGRB jet and its ejecta \citep{pmg+09,bmc+11}. Some models posit that the anomalous X-ray emission is produced by the spindown and magnetic field decay of a rapidly rotating supra-massive neutron star produced by the merger of two low-mass neutron stars \citep{rom+13,mp14,sc16,lsa19}. In fact, \citet{ptz+19} and \citet{lsl+19} have recently used X-ray evidence to argue that such a remnant may have been produced by GRB 170817A/GW 170817 \citep{aaa+17_a,aaa+17_b}. It is not known how massive neutron stars can be before collapsing to black holes \citep{dpr+10,mjw+18,cfr+19}, but some SGRB remnants could act as central engines powering X-ray emission and possibly other electromagnetic phenomena.

In 2007, an altogether different type of astrophysical transient was discovered: fast radio bursts \citep[FRBs;][]{lbm+07}. FRBs are typically several milliseconds in duration with radio emission spanning several hundred MHz and peak flux densities ranging from approximately 0.1 to 10 Jy. Their dispersion measures (DM), the integrated column density of free electrons between the source and observer, are well in excess of what is anticipated from Galactic electron density models \citep{cl02, ymw+17}, strongly suggesting extragalactic origins. The number of known FRBs grew slowly for years after the initial discovery \citep[e.g.,][]{tsb+13,sch+14, mls+15,cpk+16}, but has grown precipitously with the advent of new instruments \citep{smb+18,CHIME19_a}. Efforts to localize FRBs through interferometry are technically challenging and were initially unsuccessful, leading only to upper limits on the rate of FRBs exceeding a certain fluence \citep{lbb+15}. Currently, eleven sources are known to produce repeat FRBs \citep{ssh+16,CHIME19_b, abb+19}, indicating non-cataclysmic sources. One repeating FRB source, FRB 121102, has been interferometrically localized to a dwarf galaxy more than 900 Mpc away, confirming its extragalactic origin \citep{clw+17, tbc+17,lab+17}. Recently, two FRBs that have not yet been observed to repeat have also been localized to host galaxies at cosmological distances \citep{bdp+19,rcd+19}. See \citet{phl19} and \citet{cc19} for recent comprehensive reviews of the FRB literature.

Young extragalactic neutron stars or magnetars paired with plasma-lensing phenomena could potentially explain many observed properties of the FRB population, including the existence of repeating sources \citep{cw16,cwh+17}. As an example, the brightest giant pulse from the Crab pulsar observed by \citet{he07} would have a peak flux density in excess of 10 ${\rm \mu Jy}$ if the Crab were 1 Gpc away. There are very few pulsars known outside of the Milky Way \citep[e.g.,][]{mlk06} because standard pulsar emission is not bright enough to be seen from extragalactic distances and high DMs can smear individual radio pulses into one another, obscuring the pulsar phenomenon. Efforts to detect extragalactic neutron stars through their giant pulses predate the initial discovery of FRBs, but no such pulses were found \citep[see, e.g.,][and references therein]{mc03}.  

Motivated by the possibility that SGRBs could mark the birthplaces of extragalactic neutron stars or magnetars potentially capable of generating FRBs, we carried out an extensive campaign of targeted searches for FRBs from the sites of six low-redshift SGRBs between 2016 September and 2017 July. Since our observations, the sample of FRBs has grown substantially and there has been much development in the theoretical modeling of FRB sources. After FRB~121102 was localized and found to be coincident with a steady, non-thermal radio source in a star-forming dwarf galaxy, \citet{mbm17} argued that the properties of the persistent source and host galaxy were consistent with a remnant of a long GRB or a superluminous supernova, both classes of objects believed to be powered by millisecond magnetars. In follow-up work, \citet{mm18} developed a few-parameter model of an evolving core-collapse supernova remnant that describes many of the key observational properties of FRB 121102: a flaring, decades-old magnetar produces the FRBs and drives a wind of relativistic electrons into an expanding magnetized nebula; persistent synchrotron radiation is generated at the wind's termination shock; the exceptionally high rotation measure of the FRBs \citep{msh+18} is produced by cooler electrons injected into the nebula early in its expansion. \citet{mal+19} recently published the results of a targeted FRB search similar to the one we describe here but with only a quarter of our total observing time and focusing on long GRB remnants---they found none.

Nonetheless, \citet{mbm19} admit that binary neutron star mergers are still a feasible channel through which magnetars capable of producing FRBs may be created, and that the FRB recently localized by \citet{bdp+19} to the outskirts of a massive quiescent galaxy may be better explained by a magnetar created through a binary neutron star merger. SGRBs may be just one of several channels through which objects capable of producing FRBs are formed. 

We found no FRBs from the SGRB sites we investigated. In Section~2 we describe the SGRBs we investigated and why we selected them. In Section~3 we describe our observations. In Section~4 we describe the FRB search to which we subjected our data. In Section ~5 we discuss some of the implications of our non-detection and place upper limits on the energy of FRBs that could have gone undetected during our observations. Finally, in Section~6, we offer concluding remarks.

\section{Source Selection}
NASA's {\it Swift} satellite \citep{gcg+04} has discovered more than a thousand gamma-ray bursts (both long and short) since its launch in 2004. We made extensive use of their online catalog\footnote[1]{\url{https://swift.gsfc.nasa.gov/archive/grb\_table/}} when selecting which sources to target for our investigation. As such, five of the six SGRB sites we observed were first detected by {\it Swift}. The sixth was first detected by NASA's {\it HETE-2} satellite \citep{vvd+99}, a predecessor to {\it Swift}.

We considered only SGRBs that occurred north of approximately $-40^\circ$ declination so that they could be observed with the Green Bank Telescope (GBT). We then considered only SGRBs with measured redshifts $z\lesssim0.25$. The precise value of this redshift threshold was chosen arbitrarily, but the goal was to prioritize nearby SGRBs so that any low-luminosity FRB emission could be more readily detected. Applying this threshold dramatically shrunk the pool of potential targets. We made one exception to this proximity criterion for GRB~130603B ($z=0.356$) for reasons we discuss below.   

Since the maximum possible neutron star mass is not known, if a neutron star is produced in an SGRB, it is unclear if it will be stable for long periods of time or will quickly collapse to a black hole after a brief period of rapid spin down. It is also unclear if a young neutron star could emit the super-giant radio pulses that would appear as FRBs if it is embedded in a relatively dense cloud of ejecta from the SGRB explosion. It may take some period of time for the neutron star remnant's environment to evolve to a point conducive to FRB generation. Because of these unknowns, we did not consider how long ago an SGRB occurred when deciding which sources to search and eventually investigated sources with ages ranging from 0.1 to 12 years, spanning the whole range of ages of SGRB remnants that were precisely localized at the time of our observations. 


\begin{table*}
\centering
\begin{tabular}{lccccccp{3.5cm}}
            \hline 
            \hline 
            \noalign{\smallskip}
            {\rm Source}  &  {\rm R.A.}    &   {\rm Dec}  &   {\rm Redshift}    &   {\rm Distance}  &   {\rm DM}$_{\rm gal}$    &   {\rm Age}   &   Reference\\
                        &   {\rm (hh:mm:ss)}    &   {\rm (dd:mm:ss)}    &   &   {\rm (Gpc)} &   {\rm (pc~cm$^{-3}$)}   &   {\rm (yr)} &  \\   
            \hline
            \noalign{\smallskip}
            {\rm GRB~050509B}    &   12:36:14    &   +28:59:05      &   0.225  &   1.16 & 19.85  &   12.0 &   \citet{hrb+05} \\
            {\rm GRB~050709}     &   23:01:27    &   $-$38:58:40    &   0.160  &   0.79 & 32.87  &   11.8 &  \citet{bra+05} \\
            {\rm GRB~080905A}    &   19:10:42    &   $-$18:52:49    &   0.122  &   0.59 & 177.57  &   8.6 &    \citet{pbb+08} \\
            {\rm GRB~130603B}    &   11:28:48    &   +17:04:18      &   0.356  &   1.96 & 29.28  &   3.9 &    \citet{mbb+13} \\
            {\rm GRB~150101B}    &   12:32:05    &   $-$10:56:02    &   0.134  &   0.65 & 36.57  &   2.4 &      \citet{c15}   \\
            {\rm GRB~160821B}    &   18:39:55    &   +62:23:31      &   0.160  &   0.79 & 55.54  &   0.1 &    \citet{sbb+16} \\
            \hline
\end{tabular}
      \caption{\label{tab:sources} The sources we observed, their celestial coordinates (J2000), their redshift, inferred luminosity distance, the maximum Galactic contribution to DM along that line of sight as predicted by the NE2001 electron density model, the time between the SGRB detection and when we began our observations, and the initial SGRB discovery announcement citation.}
\end{table*}

In Table 1, for each of the six sources we targeted, we list the name, right ascension, declination, redshift, age, and reference to the discovery announcement from the GRB Coordinates Network (GCN) Circulars. Based on the redshift measurements, we have computed the luminosity distance to each SGRB (also in Table 1) assuming a flat cosmology with Hubble parameter $H_0=67.4$~km~s$^{-1}$~Mpc$^{-1}$ and matter density $\Omega_m=0.315$ \citep{Planck18}. The age is the time in years between when the SGRB occurred and when we first observed it. We also list the maximum Galactic contribution to DM, DM$_{\rm gal}$, predicted by the NE2001 model of \citet{cl02} for each line of sight. In the following subsections, we discuss each of the sources we targeted in more detail.

\subsection{GRB~050509B}
GRB~050509B was described by \citet{bpp+06} as a ``watershed event" in the study of SGRBs. In this event, the {\it Swift} Observatory, for the first time, enabled the rapid localization of an SGRB which facilitated follow-up observations with a variety of ground-based instruments beginning just eight minutes after the initial burst. The SGRB afterglow was sufficiently proximal to a bright elliptical galaxy to be confidently associated with it. The galaxy's redshift was $z=0.225$, making GRB~050509B the earliest SGRB to be conclusively shown to have originated from cosmological distances.

The afterglow of GRB~050509B was intrinsically dim in X-rays and was also obscured by diffuse X-ray emission from the galaxy cluster that the SGRB's host galaxy is in \citep{bpp+06}. Though it did not display any of the anomalous X-ray activity seen from some SGRBs that is potentially tied to a neutron star remnant, GRB~050509B has, to this day, the smallest measured redshift of any SGRB visible to the Arecibo telescope. Owing to the large collecting area of Arecibo and the opportunity to potentially detect very faint FRBs from a relatively nearbye SGRB remnant, we included GRB~050509B among our targets.

\subsection{GRB~050709}
GRB~050709 is the one object we targeted that was discovered by the {\it HETE-2} satellite rather than {\it Swift}. Rapid follow up observations of GRB050709 with the {\it Chandra} X-ray Observatory yielded a confident detection of an X-ray afterglow and identification of a host galaxy with a redshift $z=0.160$. Sixteen days after GRB~050709, \citet{ffp+05} found that the X-ray luminosity of the afterglow had only decreased by a factor of two and that it was likely flaring intermittently. They argued that this was potentially evidence for continued injection of energy into the ejecta from some persistent central engine. The low redshift of GRB~050709 paired with evidence for the presence of a central engine lasting well after the initial SGRB made it an obvious choice for our list of targets.

\vspace{.25cm}
\subsection{GRB~080905A}
Following the {\it Swift} discovery of GRB~080905A, \citet{rwl+10} conducted deep optical observations and found an optical counterpart and host galaxy for the SGRB. They measured a redshift of $z=0.1218$, the lowest ever measured for an SGRB at that time\footnote[2]{LIGO has since used gravitational wave measurements to infer a much lower redshift of $z=0.008$ for GRB~170817A/GW~170817 \citep{aaa+17_a,aaa+17_b}.}. Unfortunately, GRB~080905A showed no hints of persistent central engine activity to the point that \citet{sc16} cited it as a canonical example of just a thermally radiating expanding shell of ejecta. Nonetheless, because of its proximity to Earth among the population of known SGRBs, we targeted GRB~080905A.

\subsection{GRB~130603B}
An optical afterglow from GRB~130603B enabled a redshift determination of $z=0.356$. It is the most distant SGRB we targeted. It was notable because it was associated with a potential kilonova, a near-infrared transient fueled by the decay of heavy radioactive elements \citep{tlf+13}. But the reason we relaxed our requirement that $z\lesssim0.25$ for this source was that there was strong evidence that a magnetar remnant was produced by the merger that caused the SGRB. \citet{fbm+14} reported anomalously high X-ray emission more than a day after the SGRB. They, along with \citet{mp14}, argued that accretion onto and rapid spindown of a magnetar remnant could be generating the X-rays. Also, the issue that any FRB emission from this source would be made comparatively faint because of the large distance was ameliorated by the fact that the site of GRB~130603B is visible with Arecibo. \citet{mal+19} recently reported a targeted search for FRBs from the sites of mostly long GRBs, but they included GRB 130603B as the sole SGRB in their search. 

\subsection{GRB~150101B}
Association of GRB~150101B with an optical and X-ray afterglow allowed for the determination of a redshift $z=0.1343$. It was a compelling target for our investigation because when we were preparing our observing proposals in mid-2016, GRB~150101B was the most recent SGRB visible to the GBT with a redshift $z\lesssim0.25$. Additionally, \citet{lht+15}, based on two epochs of {\it Chandra} observations taken seven and 39 days after the initial burst, found the X-ray afterglow to be fading relatively slowly so long after the burst. We took this as possible evidence for ongoing energy injection into the afterglow from a persistent central engine. After we proposed our observations, \citet{fmc+16} carried out extensive analysis of the GRB~150101B afterglow based on observations across the electromagnetic spectrum---they did not find any compelling evidence for a persistent neutron star remnant, but could also not conclusively rule one out.     

\subsection{GRB~160821B}
Shortly after we submitted a proposal to observe five SGRB remnants with the GBT, GRB~160821B occurred. An optical afterglow was detected and a redshift of $z=0.16$ was determined for the host galaxy \citep{xmd+16,lwt+16}. Among SGRBs with redshift measurements, GRB~160821B was the third closest to ever occur and be visible to the GBT. Based on its proximity and recency alone, we requested Director's Discretionary Time with the GBT to search GRB~160821B for FRBs. All of our observations of this source occurred between 40 and 47 days after the SGRB. Subsequent analysis by \citet{lzz+17}, \citet{tcb+19}, and \citet{ltl+19} found that various features of the X-ray and near infrared afterglow of GRB~160821B could be explained by the presence of a neutron star remnant.

\section{Observations}

 The first detection of an FRB followed by the next several detections with the Parkes radio telescope were done with radio frequencies between 1.2 and 1.5 GHz \citep{lbm+07,tsb+13}. The recent FRB detections from the \citet{CHIME19_a, CHIME19_b} were made with radio frequencies between 400 and 800 MHz, the lowest-frequency FRB detections to date. The first repeating FRB source has shown bursts with complex frequency-dependent structure, sometimes decreasing in brightness with increasing radio frequency, sometimes vice versa \citep{ssh+16,hss+19}, and the bursts have been detected from 400 MHz to 8.4 GHz \citep{clw+17,gsp+18,jcf+19}.

All of this is to say that there is not currently enough known about FRBs to justify using a particular part of the radio band to conduct a search over another. With the GBT, we searched between 1.6 and 2.4 GHz (S-band). Since the two sources we observed with Arecibo were also observed with the GBT, to increase our frequency coverage, we chose to search a different part of the band with Arecibo: 980 to 1780 MHz (L-band). We used the PUPPI and GUPPI backends at Arecibo and Green Bank, respectively---they are identical \citep{drd+08}. We used 2,048 frequency channels, the best possible frequency resolution available with PUPPI and GUPPI. This minimizes dispersive smearing across channels for highly dispersed narrow pulses. We sampled every 40.96 ${\rm \mu s}$, the fastest possible sampling rate with PUPPI and GUPPI given our choice of bandwidth and frequency resolution. 

For a flat-spectrum, temporally resolved, boxcar shaped pulse of width $W$, the minimally detectable flux density is \begin{equation}
S_{\rm min}=\frac{T_{\rm sys}~({\rm S/N})_{\rm thresh}}{G(2WB)^{1/2}},    
\end{equation}
where $T_{\rm sys}$ is the system temperature, $({\rm S/N})_{\rm thresh}$ is the signal-to-noise threshold required to claim a detection, $G$ is the telescope gain, and $B$ is the system bandwidth. At both Arecibo and the GBT, we utilized 800~MHz of bandwidth. The gain of the GBT at S-band is $\approx$~2~K~Jy$^{-1}$ and the system temperature is  $\approx$~22~K. The gain of Arecibo at L-band is $\approx$~10~K~Jy$^{-1}$ and the system temperature is $\approx$~27~K. With these observing parameters fixed, the minimum detectable peak flux density is
\begin{equation}
S_{\rm min}\approx A_\star~\left(\frac{({\rm S/N})_{\rm min}}{10}\right)\left(\frac{W}{1~{\rm ms}}\right)^{-1/2},
\end{equation}
where $A_\star=87$ mJy for the GBT and $A_\star=21$ mJy for Arecibo. Staying with a fiducial burst width of 1 ms, the minimally detectable burst fluence in our GBT data is 0.087~Jy~ms and for our Arecibo data, it is 0.021~Jy~ms. In Tables 1 and 2 from \citet{cc19}, they collate the minimally detectable burst fluence for every major FRB survey done to date. The sensitivity of our GBT data is surpassed only by previous work done with Arecibo and our Arecibo data is more sensitive than any other studies carried out at Arecibo by about a factor of two. Any real FRB will have more temporal and spectral structure than the idealized pulse we have considered here, so these should be taken as mildly optimistic measures of our sensitivity.

\citet{pjk+15}, with between nine and 33 hours of follow-up observations, found no repeat bursts from the sites of eight FRBs detected with the Parkes radio telescope. Nearly 100 hours of follow-up observations of the first FRB reported by \citet{lbm+07} failed to detect any repeat bursts. But, in the case of the first repeating FRB, \citet{ssh+16} discovered 10 bursts in three hours of follow-up observations with Arecibo. Our hope with these observations was to find evidence of FRBs from sources capable of producing repeating FRBs. Based on these earlier examples, we decided that 10 hours of observations per target would give us a good chance of detecting FRB emission if it was occurring. As further support for 10 hours of integration time per target being a reasonable amount, after our observations took place, the \citet{CHIME19_a} discovered a second repeating FRB. They detected six repeat bursts in approximately 23 hours of integration time. 

\section{Search Procedure and Results} 

\begin{figure}
\begin{center}
\hspace{-5mm}
\includegraphics[scale=.36]{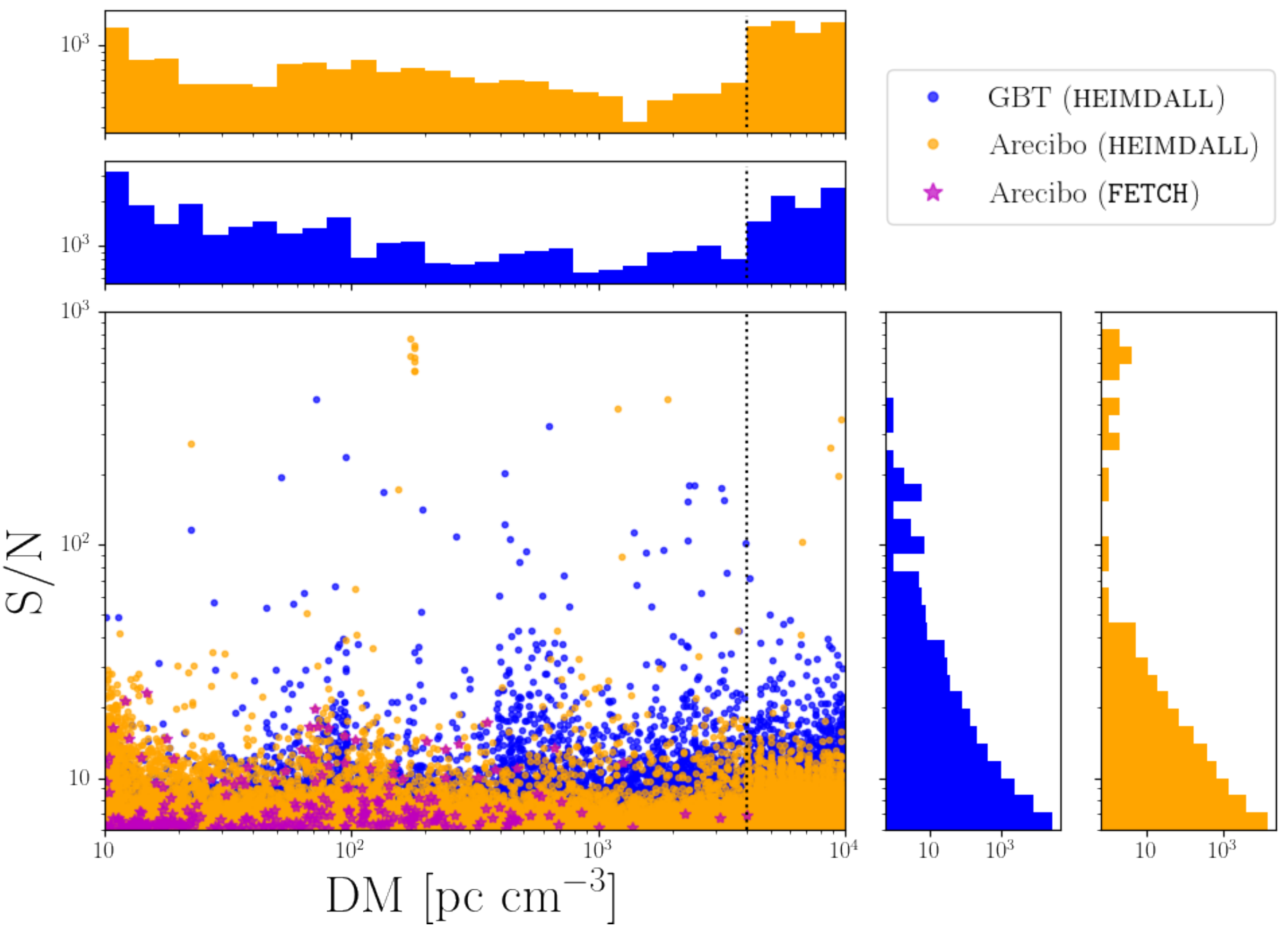}
\caption{\label{fig:scatter} The distribution of FRB candidates in the S/N-vs.-DM plane from the \textsc{heimdall} search of our data and the much smaller population of Arecibo candidates \texttt{FETCH} flagged for further visual inspection (magenta stars). The orange (blue) dots and histograms correspond to the \textsc{heimdall} candidates from Arecibo (GBT) data. The vertical dotted black line indicates that we had to decimate the resolution of our data to extend our search to high DMs between 4,000 and 10,000 pc cm$^{-3}$.}
\end{center}
\end{figure}

\begin{figure}
    \centering
    \includegraphics[width=\columnwidth]{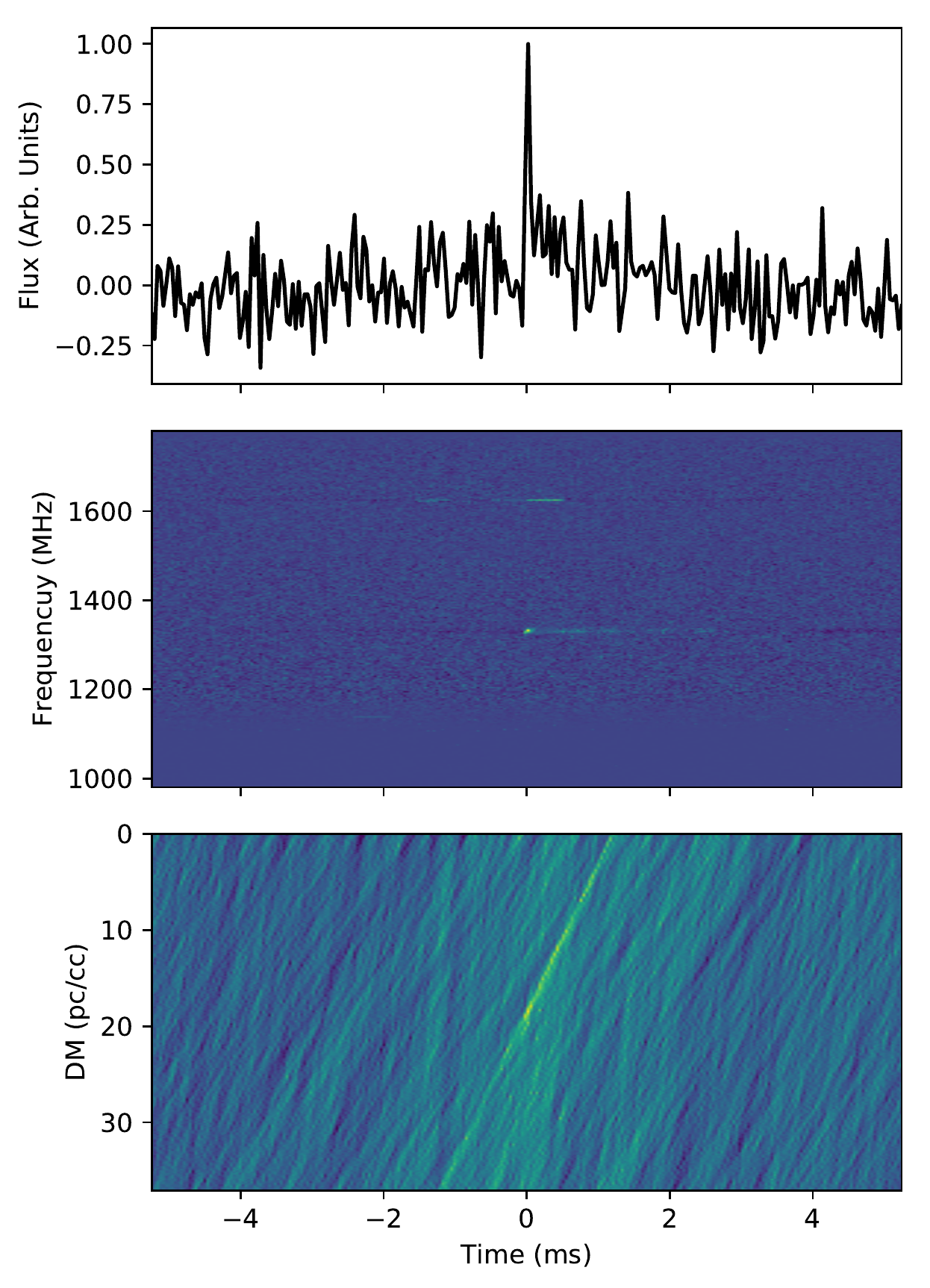}
    \caption{An example of RFI in Arecibo data that \texttt{FETCH} erroneously classified as a legitimate FRB. \texttt{FETCH} works by looking for characteristic patterns in images such as this. {\bf Top:} the flux-density summed across the band as a function of time where the band has been dedispersed to optimize the S/N of the \textsc{heimdall} candidate. {\bf Middle:} the dynamic spectrum, or the flux density in the frequency-vs.-time plane, dedispersed according to the optimal DM. This panel shows that this candidate was clearly two bursts of narrow-band RFI, the lower frequency burst delayed slightly relative to the higher frequency burst.  {\bf Bottom:} How the S/N varies as the trial DM and centroid of the boxcar template are varied. This panel appears approximately as it should for a legitimate FRB. \texttt{FETCH} will become more discriminating with further training on a broad variety of data.}
    \label{fig:RFIcand}
\end{figure}

To search for FRBs, we employed \textsc{heimdall}\footnote[3]{\url{https://sourceforge.net/projects/heimdall-astro/}}, a GPU-accelerated algorithm for fast incoherent dedispersion and boxcar convolution \citep{bbb+12}. We searched our full-resolution data for FRBs with durations between one and 1,024 time samples (with increments in powers of two) and with DMs between 0 and 4,000 pc cm$^{-3}$. We extended our search to DMs as high as 10,000 pc cm$^{-3}$, but due to limitations in our GPU hardware, in order to extend our search above DM values of 4,000~pc~cm$^{-3}$, we had to decimate the data by a factor of two in frequency and five in time before inputting it to \textsc{heimdall}. The effective resolution of this decimated data was approximately 0.78~MHz in frequency and 204.8~$\mu$s in time. On the low end of our trial DM range, the values are smaller than the anticipated Galactic contribution. However, considering low DM trials is useful for diagnosing radio frequency interference (RFI). No FRBs have been found with a DM in excess of 3,000 pc cm$^{-3}$, but with little understanding of the extent to which local material could enhance the DM of FRBs from an SGRB, we extended our search to much higher trial values.

The \textsc{heimdall} search resulted in 59,815 candidates above a S/N of six, 37,999 in GBT data and 21,816 in Arecibo data. The distribution of all candidates in the S/N-vs.-DM plane is shown in Figure \ref{fig:scatter}. Their density falls off steeply with increasing S/N for both the Arecibo and GBT data. The search of our decimated data above 4,000 pc~cm$^{-3}$ had a higher density of trial DMs per logarithmic interval than our search of the full-resolution data, so the distribution of \textsc{heimdall} candidates in DM has a clear uptick above 4,000 pc~cm$^{-3}$. 

The \textsc{heimdall} candidates were further classified using \texttt{FETCH}\footnote[4]{\url{https://github.com/devanshkv/fetch}} to help distinguish between RFI and FRBs \citep[see][for full \texttt{FETCH} implementation details]{aab+19}. \texttt{FETCH} works by looking for characteristic patterns in images like Figure~2 generated for each of the \textsc{heimdall} candidates. We found that strong, impulsive RFI occasionally caused a small patch of pixels in the dynamic spectrum (dedispersed intensity on a frequency-vs.-time plane) to take on such large values that all other values would be driven to their lowest possible value. With such a degraded dynamic range in the dynamic spectrum, \texttt{FETCH} was incorrectly identifying these as legitimate candidates. To prevent this, we clipped any values in the dynamic spectrum that were more than 10 standard deviations above the mean. With this clipping implemented,  \texttt{FETCH} labelled 259 candidates as potential FRBs, all of which were then visually inspected and found to be false positives. 

All candidates that \texttt{FETCH} flagged for visual inspection were in our Arecibo data. The neural network at the heart of \texttt{FETCH} was trained on an abundance of GBT data, so \texttt{FETCH} is adept at discriminating against false positives created in the RFI environment of Green Bank. Figure~2 shows an example of one recurring variety of RFI-induced false positive at Arecibo. Two narrow-band impulses of RFI near 1.3 GHz and 1.6 GHz occur with just such a delay between them that when the frequency channels are shifted in accordance with a trial DM of approximately 20~pc~cm$^{-3}$ the two impulses appear roughly coincident. The dynamic spectrum (the middle panel of Figure \ref{fig:RFIcand}) is clearly spurious, but the S/N of the detection in the DM-vs.-time plane (the bottom panel of Figure \ref{fig:RFIcand}) resembles that of a legitimate FRB. With additional training on Arecibo data, future iterations of \texttt{FETCH} will be able to readily recognize this and other varieties of RFI-induced false positives.

\section{Implications of Non-detection}

Recently, in reporting a non-detection of FRBs from a targeted search of mostly long GRBs (but also GRB 130603B), \citet{mal+19} placed upper limits on the rates of FRBs from the sources they searched. All of their constraints were predicated on the assumption that FRBs from repeating sources are generated as Poisson processes. While this is a convenient assumption to make, it does not describe FRB 121102, the only thoroughly studied repeating FRB source. FRB 121102 appears to go through episodic outbursts followed by long spans of inactivity. We are currently working on ways to model and constrain this type of intermittency, but reserve any such discussion to future work. The sites we investigated may simply not host objects capable of producing FRBs or may be going through prolonged periods of quiescence.

Since the distances to all of the SGRBs in our sample are known, we can place upper limits on the energy of any FRBs that may have been generated at the sources during our observations. As we discussed in Section 3, our fluence limit, $F_{\rm min}$, for a 1~ms fiducial pulse width and a S/N threshold of 10 are 0.087~Jy~ms and 0.021~Jy~ms at the GBT and Arecibo, respectively. Assuming the radio emission from any bursts filled a band of radio frequencies $\Delta\nu$ at the telescope and that the emission was beamed into a fraction $f_b$ of the sphere, the maximum energy that could have been emitted for a source at luminosity distance $D$ is 
\begin{eqnarray}
E_{\rm max}=4\pi f_b D^2(1+z)\Delta\nu F_{\rm min}
\end{eqnarray}
\citep[for a discussion of this formula, see, e.g.,][]{cc19}. The redshift enters this expression to convert the bandwidth observed at the telescope to the emission bandwidth at the source. Though FRB emission is possibly tightly beamed, we assume isotropic emission, i.e. $f_b=1$, since there is no theoretically or observationally well-motivated smaller value and it will facilitate comparison to other work in the literature in which this assumption was also made. We assume that $\Delta\nu=B$, our full 800~MHz bandwidth. Many FRBs span smaller bands than this, but such wideband emission is not unprecedented---some bursts from FRB 121102 have been seen to span more than 1 GHz.

\begin{table}
\centering
\begin{tabular}{lcc}
            \hline 
            \hline 
            \noalign{\smallskip}
            {\rm Source}    &   {\rm Telescope}  &  {$E_{\rm max}/10^{38}$}\\
                            &                   &   {\rm (erg)}\\ 
            \hline
            \noalign{\smallskip}
            {\rm GRB~050509B}   &   GBT     &   1.38  \\
            {\rm GRB~050509B}   &   Arecibo &   0.33  \\
            {\rm GRB~050709}    &   GBT     &   0.60  \\
            {\rm GRB~080905A}   &   GBT     &   0.33  \\
            {\rm GRB~130603B}   &   GBT     &   4.35  \\
            {\rm GRB~130603B}   &   Arecibo &   1.05  \\
            {\rm GRB~150101B}   &   GBT     &   0.40  \\
            {\rm GRB~160821B}   &   GBT     &   0.60  \\
            \hline
\end{tabular}
      \caption{\label{tab:sources} The maximum energy an FRB from any of our sources could have had during our observations without being detected. We have conservatively assumed isotropic emission spanning our full 800~MHz instrumental bandwidth. The burst duration is assumed to be 1~ms and we are assuming a S/N detection threshold of 10. }
\end{table}

Using the luminosity distances from Table 1, we have computed $E_{\rm max}$ for all of the SGRBs in our sample and compiled them in Table 2. Our loosest constraint is $E_{\rm max}\approx4.35\times10^{38}$ erg from our GBT observations of GRB 130603B, which has a luminosity distance of nearly 2 Gpc, almost twice as distant as any other SGRB in our sample. Our tightest constraint is $E_{\rm max}\approx3.3\times10^{37}$ erg from our GBT observations of GRB 080905A and our Arecibo observations of GRB 050509B. To put these energy constraints in context, consider the study of low-energy bursts from FRB 121102 done by \citet{gms+19}. In just over 3 hours of 1.4-GHz observations with Arecibo during an eruption of activity from FRB 121102, they detected 41 bursts with isotropic-equivalent energies ranging from approximately $10^{37}$ erg to approximately $2\times10^{38}$ erg. Had any FRBs with as much energy as the most energetic of the bursts detected by \citet{gms+19} occurred during our observations, we would have detected them unless they occurred during our GBT observations of GRB 130603B, our most distant target. If we relax our strict S/N threshold of 10 by a factor of approximately 2, we would have detected some of the flaring activity seen by \citep{gms+19} even from this most distant site we investigated.       

\section{Conclusions}
Detecting an FRB in this search would have connected FRBs to SGRB remnants and shown that SGRBs can produce long-lived massive neutron stars. But our non-detection does not prove the lack of such connections. Stable central engines may only be produced in a small fraction of SGRBs. We targeted multiple SGRBs with X-ray indications of persistent energy injection from a central engine, but there may have been no magnetars at all at some or all of the sites we searched. 

Even if some of the sites we targeted are capable of generating FRBs, they may be so intrinsically faint as to be difficult to detect or the emission was band-limited and unfortunately outside of our range of observing frequencies. With our use of wideband instruments at two of the largest radio telescopes in the world, our sensitivity is difficult to surpass without the use of new instruments---the ultra-wideband receivers now in use at the Parkes Radio Telescope \citep{dbb+15} and under development for the GBT, for example---or the still larger Five-hundred-meter Aperture Spherical Telescope \citep[FAST;][]{lp16}. Furthermore, any FRBs generated from the sites we targeted may be so intermittent that much more than 10 hours of observation would be required to detect a single burst. With only a small number of FRB sources known to repeat, no robust inferences about intermittency can be established at this time. Magnetars produced through the SGRB channel will be enshrouded in less dense nebulae of ejecta than those produced by core-collapse supernovae \citep{mbm19}. If magnetar winds interacting with shells of ejecta are producing FRBs, more diffuse ejecta may lead to greater intermittency in FRB production. That FRBs from magnetars produced by neutron star mergers may be extremely intermittent compared to FRB 121102, which sometimes produces multiple bursts within minutes and sometimes becomes quiescent for months, decreases the likelihood of success for a search such as the one we have conducted, regardless of instrumental sensitivity. 

If the connection between SGRBs and FRBs exists, it will likely be made in the next few years by instruments such as the Australian Square Kilometer Array Pathfinder (ASKAP) or the Deep Synoptic Array (DSA), arrays of radio telescopes capable of surveying tens of square degrees of the sky at once and localizing any FRBs with arcsecond precision \citep{bdp+19,rcd+19}. In the course of detecting potentially hundreds of FRBs each year, some may be found to be spatially coincident with the sites of known SGRBs.

\vspace{0.5cm}
\emph{Author contributions}: The initial observing proposals were conceived of and developed by D.R.M. with substantial input and assistance from S.C., J.M.C., D.R.L., M.A.M., S.M.R., and R.S.W. Observations were carried out by H.T.C., M.T.L., D.R.L., R.S.L., D.R.M., and S.M.R. Processing of the data was done by D.A., K.A., N.G.D., D.R.M., and O.Y. Development of this manuscript was lead by D.R.M. with contributions and input from all co-authors.

\begin{acknowledgments}
The Green Bank Observatory is a facility of the National Science Foundation (NSF) operated under cooperative agreement by Associated Universities, Inc. (AUI). 
The Arecibo Observatory is operated by the University of Central Florida, Ana G. M\'{e}ndez-Universidad Metropolitana, and Yang Enterprises under a cooperative agreement with the NSF (AST-1744119).
At the outset of this project, D.R.M. was a Jansky Fellow of the National Radio Astronomy Observatory (NRAO). NRAO is a facility of the NSF operated under cooperative agreement by AUI. Several authors are members of the NANOGrav collaboration, which receives support from NSF Physics Frontiers Center award number 1430284. K.A. acknowledges support from NSF grant AAG-1714897. M.A.M., D.A., and D.R.L. have additional support from NSF OIA-1458952 and NSF AAG-1616042. S.C. and J.M.C. acknowledge support from NSF AAG-1815242. D.R.L. acknowledges support from the Research Corporation for Scientific Advancement. R.S.W. acknowledges financial support from the European Research Council (ERC) for the ERC Synergy Grant BlackHoleCam under contract no. 610058.
\end{acknowledgments}

\bibliographystyle{aasjournal}
\bibliography{trans_GRB_AJ}
\end{document}